\documentclass[12pt]{article}
\pdfoutput=1
\usepackage{jheppub}
\pdfoutput = 1

\usepackage{amsmath}

\usepackage{amssymb}
\usepackage{bbm}
\usepackage{bbold}
\usepackage{braket}
\usepackage{caption}
\usepackage{color}
    \definecolor{darkgreen}{rgb}{0,0.5,0}
    \definecolor{darkred}{rgb}{0.5,0,0}
    \definecolor{darkblue}{rgb}{0,0,0.6}
    \definecolor{purple}{rgb}{0.4,.2,0.7}

\usepackage[mathcal]{eucal}
\usepackage{float}
\usepackage{graphicx}
\usepackage{hyperref}

\usepackage{mathabx}
\usepackage{mathtools}
\usepackage{pdfsync}
\usepackage{slashed}
\usepackage[normalem]{ulem}
\usepackage{upgreek}
\usepackage{url}

\usepackage[ragged]{footmisc}
\def\be{\begin{equation}}
\def\ee{\end{equation}}

\renewcommand{\tilde}{\widetilde}

\numberwithin{equation}{section}

\begin{document}

%\subheader{empty}
\title{Gravitational thermodynamics without the conformal factor problem: Partition functions and Euclidean saddles from Lorentzian Path Integrals}
\author{Donald~Marolf,}

%affilations
\affiliation{Department of Physics, University of California at Santa Barbara, Santa Barbara, CA 93106, U.S.A.}
%\affiliation[b]{Department of Applied Mathematics and Theoretical Physics, University of Cambridge, Wilberforce Road, Cambridge, CB3 0WA, UK}

% e-mail addresses
\emailAdd{marolf@ucsb.edu}
%\emailAdd{jss55@cam.ac.uk}

\abstract{Thermal partition functions for gravitational systems have traditionally been studied using Euclidean path integrals.  But in Euclidean signature the gravitational action suffers from the conformal factor problem, which renders the action unbounded below.  This makes it difficult to take the Euclidean formulation as fundamental.  However, despite their familiar association with periodic imaginary time,  thermal gravitational partition functions can also be described by real-time path integrals over contours defined by real Lorentzian metrics.  The one caveat is that we should allow certain codimension-2 singularities analogous to the familiar Euclidean conical singularities.  With this understanding, we show that the usual Euclidean-signature black holes (or their complex rotating analogues) define saddle points for the real-time path integrals that compute our partition functions.  Furthermore, when the black holes have positive specific heat, we provide evidence that a codimension-2 subcontour of our real Lorentz-signature contour of integration can be deformed so as to show that these black holes saddles contribute with non-zero weight to the semiclassical limit, and that the same is then true of the remaining two integrals.}

\maketitle

%%%%%%%%%%%%%%%%%%%%%%%%%
\section{Introduction}
%%%%%%%%%%%%%%%%%%%%%%%%%

In non-gravitational field theories, thermal partition functions $Z(\beta)$ are naturally described by Euclidean path integrals.  And as pointed out long ago by
Gibbons and Hawking \cite{Gibbons:1976ue}, there is a sense in which this remains true for gravitational theories as well.  In particular, such integrals can often be evaluated in the semiclassical approximation using saddle points associated with Euclidean black holes.

Unfortunately, due to the conformal factor problem, the Euclidean gravitational action is unbounded below.  This prevents one from taking the integral over all real Euclidean metrics as a definition of the problem to be studied.  Many authors simply follow \cite{Gibbons:1978ac} and choose to integrate over a contour for which the Euclidean metrics are not real.  But while this often gives physically satisfying results (see e.g. \cite{Allen:1984bp,Prestidge:1999uq,Kol:2006ga,Headrick:2006ti,Monteiro:2008wr,Monteiro:2009tc,Monteiro:2009ke,Anninos:2012ft,Benjamin:2020mfz,Cotler:2019nbi,Marolf:2018ldl,Cotler:2021cqa}), the choice of contour is an ad hoc recipe that lacks a justification from first principles.

In contrast, various works \cite{Hartle:2020glw,Schleich:1987fm,Mazur:1989by,Giddings:1989ny,Giddings:1990yj,Marolf:1996gb,Dasgupta:2001ue,Ambjorn:2002gr,Feldbrugge:2017kzv,Feldbrugge:2017fcc,Feldbrugge:2017mbc} have argued that the fundamental definition should instead be made in Lorentz signature,  with the contour of integration taken to be defined by real Lorentz-signature metrics.  The idea is then that careful study might show the extent to which the contour can be deformed into the complex plane to yield an equivalent ``Euclidean'' path integral that comes equipped with a specific contour of integration. In the semiclassical approximation, one could then check any given saddle against this contour to determine the weight (if any) with which it contributes.

A prime advantage of this suggestion is that is it not obviously ruled out. The Lorentzian gravitational action $S$ is purely real for smooth real Lorentz-signature metrics, so that $e^{iS}$ is a pure phase.  Thus the integrand of our path integral is naturally oscillatory rather than diverging in absolute value.

While oscillatory integrals can be subtle, they often converge when treated as distributions.  Perhaps the best-known example of this feature is the representation of the Dirac delta-function as an integral over oscillatory exponentials,
\begin{equation}
\int dx \ e^{ikx} = 2\pi \delta(k).
\end{equation}
It is thus natural to suppose that the Lorentz-signature gravitational path integral should be understood in a similar manner.

The goal of this work is to argue that this approach can indeed be used to define the familiar thermal partition functions $Z(\beta)$ for gravitational systems.  Furthermore, despite the Lorentz signature starting point, we will provide evidence Euclidean-signature black hole solutions with positive specific heat provide saddle points that contribute with non-zero weight to the semiclassical approximation of our partition function.  In particular,  we will take care to show that a codimension-2 subcontour of the original contour of integration (defined by real Lorentz-signature metrics) can be deformed to access such saddles in a useful way, and that the same is then true of the remaining two integrals.

To briefly explain our setup, let us recall that the gravitational partition function $Z(\beta)$ has historically been defined as a path integral over some class of metrics with Euclidean-signature boundaries $S^1 \times Y$ where the $S^1$ has proper length\footnote{More generally, $\beta$ is some rescaled version of the proper length where the rescaling might depend on the location on the factor $Y$.  But this generalization has no effect on the argument below, so we use the above language for simplicity.  The analysis is similarly insensitive to whether $S^1 \times Y$ is an asymptotic boundary or whether lies at finite distance as in the idealized description of a reflecting cavity wall. } $\beta$. But we will instead use a natural definition of $Z(\beta)$ as an integral over a one-parameter family of {\it Lorentz}-signature path integrals $Z_L(T)$,
\begin{equation}
\label{eq:betaT}
Z(\beta) = \int dT f_\beta(T) \  Z_L(T).
\end{equation}
In particular, using $S$ to denote the Lorentz-signature gravitational action, $Z_L(T)$ is the integral of $e^{iS}$ over real Lorentz-signature metrics with boundaries $S^1 \times Y$ for which $T$ is the proper time around the $S^1$.  Further details of such path integrals and the class of metrics over which we integrate will be specified below.

Note that the analogous reformulation would be trivial for a stable non-gravitational system.  Suppose in particular that our system is defined in Lorentz signature on ${\mathbb R} \times Y$, where ${\mathbb R}$ is the time direction.  If the Hamiltonian is bounded below, then for $\beta >0$ we seek an expression of the form
\begin{equation}
\label{eq:EfromL}
{\rm Tr} \, e^{-\beta H}  = \int dT f_\beta(T) \  {\rm Tr} \, e^{-i HT}.
\end{equation}
Now, for familiar systems with an infinite number of states the trace ${\rm Tr} \, e^{-i HT}$ will typically fail to converge for any given value of $T$.  However, treating this object as a distribution in $T$ and integrating against functions $f_\beta(T)$ gives
\begin{equation}
\label{eq:EfromL2}
\int dT f_\beta(T) \  {\rm Tr} \, e^{-i HT} = {\rm Tr} \, \tilde f_\beta(H),
\end{equation}
where $\tilde f_\beta(\omega) = \int dT f_\beta(T) \  \, e^{-i \omega T}$ is just the appropriately-normalized Fourier transform of $f_\beta(T).$  For suitable functions the operator $\tilde f_\beta(H)$ is in fact trace-class and the right-hand side of \eqref{eq:EfromL2} will be well-defined.  Furthermore, since the spectrum of $H$ is bounded below by some ground-state energy $E_0$, we may obtain the canonical partition function by taking $f_\beta(T)$ to be the Fourier transform of some function $e^{-\beta \omega} g_{E_0}(\omega)$ where $g_{E_0}=1$ for $E \ge E_0$ and $g(\omega) \rightarrow 0$ sufficiently rapidly as $\omega \rightarrow -\infty$ that the Fourier transform exists.

We will make the same choice of $f_\beta(T)$ studying the gravitational partition function defined by \eqref{eq:betaT}.   In that context, a key question will be what precise definition we choose for the Lorentzian path integral that computes $Z_L(T)$.  After all, if the path integral is supported on {\it smooth} real Lorentz-signature bulk spacetimes, then nowhere in that support can the $S^1$ factor be contracted to a point in the bulk while remaining timelike.  This makes it hard to imagine how such a path integral can give rise to saddles described by the familiar Euclidean black holes (where the $S^1$ orbits of the `Euclidean time' Killing field do indeed contract to a point), even after possible deformations of the contour in the complex plane.

This tension will be resolved by including Lorentz-signature spacetimes that have certain codimension-2 singularities in the original contour of integration.    We will define precisely which singularities we allow in section \ref{sec:prelim} below by taking the space of such geometries to be closed under certain cut-and-paste operations.  As a result, it is natural to refer to them as conical singularities.

The action $S$ can then be defined on such geometries following \cite{Louko:1995jw,Neiman:2013ap,Colin-Ellerin:2020mva}.  Effectively, this reduces to using a complexified version of the two-dimensional Gauss-Bonnet theorem.  In particular, as in \cite{Louko:1995jw,Neiman:2013ap,Colin-Ellerin:2020mva}, it turns out that such singularities give imaginary contributions to the Lorentz-signature action $S$, so the path-integral integrand $e^{iS}$ is no longer just a phase.  In the $n > 1$ Renyi-entropy calculations of \cite{Colin-Ellerin:2020mva,Colin-Ellerin:2021jev} this effect suppressed contributions from the most natural geometries in which the area $A[\gamma]$ of the conical singularity $\gamma$ was large.  In contrast, in the present context we will find that the analogous computation turns out to {\it enhance} the contribution of geometries with large $A[\gamma]$ by a factor of $e^{A[\gamma]/4G}$.  In our thermodynamic context, this naturally corresponds to the fact that contributions to any partition function from a macrostate with entropy ${\sf S}$ are accompanied by  a factor of $e^{\sf S}$.

The astute reader will note that, after the inclusion of such conical singularities, the integrand of our Lorentz-signature path integral is no longer a pure phase.  As a result, this inclusion has now destroyed the very property of the Lorentz-signature path integral that was touted above as a way to avoid the conformal factor problem that plagues the Euclidean formulation.  In particular, if the magnitude of the integrand is $e^{A[\gamma]/4G}$, then integrating over the area $A[\gamma]$ of the conical singularity would appear to give a divergent result.  The key result below is thus that this is not necessarily the correct conclusion.  Indeed, we will proceed by first holding fixed the area $A[\gamma]$ and performing the rest of the path integral.  In this step, the integral is again strictly oscillatory and so can plausibly be convergent in the sense of distributions.  The output of this step is clearly an additional factor that also depends on $A[\gamma]$.   We use the stationary phase approximation to argue that this additional factor suppresses contributions from large $A[\gamma]$ enough to render the $A[\gamma]$ integral convergent.  This in itself should not be a surprise, as it directly related to the fact that the partition function also contains a factor of $e^{-\beta E}$, and that for a given class of black hole the on-shell energy $E$ is a function of $A[\gamma]$ (and perhaps a few other variables).  See section \ref{sec:disc} for brief comments both on going beyond the saddle-point approximation and on performing the integrals in other orders.

We begin by specifying the details of our Lorentz-signature path integral in section \ref{sec:prelim} below. Here and in most of this work we restrict discussion to Einstein-Hilbert gravity with cosmological constant $\Lambda$ (which may be zero), minimally coupled to some set of matter fields.  Section \ref{sec:thimbles} then briefly review some facts from the literature on saddle-point methods and makes a small-but-useful observation.  This sets the stage for a study the canonical ensemble for gravity in section \ref{sec:main} using the gravitational analogue of \eqref{eq:EfromL}. The main argument is given in section \ref{sec:submain}, with certain details relegated to appendix \ref{sec:FAS}. As described in subsection \ref{sec:rot}, for rotating and charged black holes the same argument applies in the grand canonical ensemble with fixed temperature, angular velocity, and electric potential.  The small changes required to study other ensembles are described in section \ref{sec:micro}, and sections \ref{sec:extreme} and \ref{sec:hd} comment on extreme limits and higher derivative corrections. Section \ref{sec:disc} concludes with a discussion of open issues and future directions.

\section{What spacetimes contribute to the real Lorentz-signature gravitational path integral?}
\label{sec:prelim}

Understanding the proper domain of integration for any path integral can be a deep and subtle question.  For the Wiener measure relevant to the quantum mechanics of harmonic oscillators, the domain of integration is supported on histories that are continuous but not differentiable (and more precisely on those that are H\"older continuous with index $\alpha < 1/2$).   This case may be thought of as a field theory in $0+1$ dimensions, and the support is expected to become even more singular for higher dimensional theories.  This is a direct analogue of the well-known fact that UV divergences of perturbative QFT also grow in strength with increasing spacetime dimension.

In general, one expects the support of the path integral measure to be determined by the action.  For gravity we must then consider the Einstein-Hilbert action which is not only complicated, but also perturbatively non-renormalizable.  A full and proper treatment of singular geometries may thus depend on the details of the ultraviolet completion of the theory.

Here we will merely propose that this support may be taken to include spacetimes with a certain class of codimension-2 singularities on which the Einstein-Hilbert action is naturally regarded as being finite, and for which higher derivative corrections to this action (which arise from e.g. including loop effects order-by-order in perturbation theory) can plausibly be regarded as being small after appropriate renormalizations.    The latter point suggests that, despite first appearances, the inclusion of such singularities does not introduce strong sensitivity to the ultraviolet completion of our theory.

We refer to this class of singularities as Lorentz-signature conical defects, and we describe them in more detail below.  The discussion here is a slight generalization of that of \cite{Colin-Ellerin:2020mva}, which was in turn inspired by \cite{Louko:1995jw} and \cite{Neiman:2013ap}.

Euclidean signature conical singularities have been shown to satisfy an analogous list of properties. In that case it is manifest that the Einstein-Hilbert action is finite, as the Ricci scalar is a Dirac delta-function while the metric is continuous, so that the integral of $\sqrt{g}R$ is well-defined. And despite the large curvature at the singularity, it was shown in \cite{Dong:2013qoa} that using conical singularities to compute higher-derivative corrections to gravitational entropy nevertheless gave corrections that were perturbatively small.  Furthermore, while the conical singularity leads to divergences in the naive higher-derivative action, the work of \cite{Dong:2019piw} showed that such divergences can be cancelled by counter-terms.  This allows one to define a finite action where all higher-derivative corrections near conical singularities remain perturbatively small.

More recently, it was noticed in \cite{Colin-Ellerin:2020mva} that a related class of singularities for Lorentz-signature metrics could be defined by starting with a collection of smooth Lorentz-signature spacetimes and performing certain cut-and-paste operations; see also comments in \cite{Marolf:2020rpm,Marolf:2021ghr}.  The idea was to follow \cite{Louko:1995jw} and \cite{Neiman:2013ap} in using the complex version of the Gauss-Bonnet theorem to define the Einstein-Hilbert action on such spacetimes, and to treat higher-derivative corrections via a formal analytic continuation of the Euclidean power series analysis performed in \cite{Dong:2019piw}.  More will be said about this last step in \cite{highDkink}. However, in order to focus on the Einstein-Hilbert case that is of primary interest, we will postpone further comments on higher-derivative corrections to section \ref{sec:hd}, where we can explain in more detail how they can be perturbatively incorporated into the main argument of section \ref{sec:main}.

We will now present a slightl generalization of the Einstein-Hilbert analysis of \cite{Colin-Ellerin:2020mva}.  In doing so, let us suppose that we are given $n$ smooth Lorentz-signature spacetimes ${\cal M}_i$ (for $i=1,\dots n$) and a smooth codimension-2 spacelike surface $\gamma$ (without boundary) in ${\cal M}_1$. We also suppose that this same $\gamma$ can be smoothly embedded as a codimension-2 surface into each of the other spacetimes ${\cal M}_i$ so that it has the same induced geometry in each ${\cal M}_i$.  For each $i$, let $\psi_i: \gamma_1 \rightarrow \gamma_i$ be a fixed metric-preserving diffeomorphism between $\gamma_1=\gamma$ in ${\cal M}_1$ and the copy $\gamma_i$ of $\gamma$ in ${\cal M}_i$, where we choose $\psi_1$ to be the identity.  We note that these $\psi_i$ are unique up to symmetries of $\gamma$.

\begin{figure}[h!]
    \centering
    \vspace{1cm}
    \includegraphics[width=0.5\textwidth]{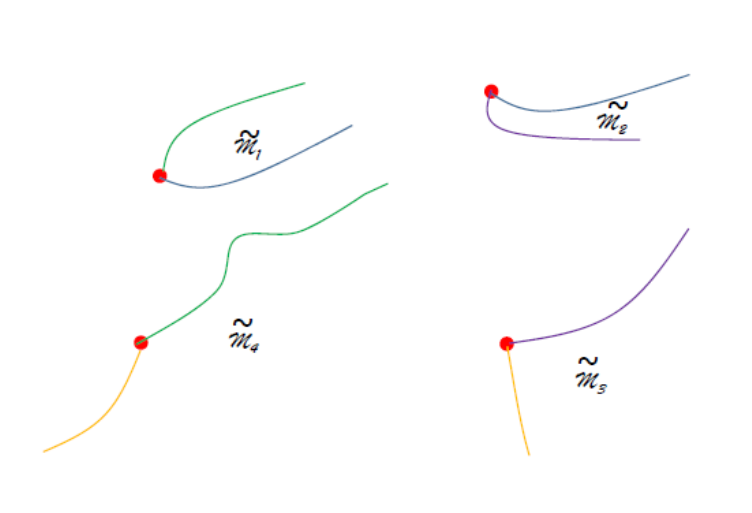}    \includegraphics[width=0.25\textwidth]{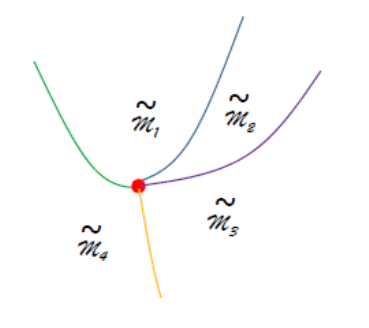}
    \caption{\label{fig:cutandpaste} An example of our cut-and-paste construction, where we cut pieces $\tilde {\cal M}_i$ from four smooth spacetimes (left) and then paste them together cyclicly (right) to form a new spacetime which may include a conical singularity.  Each piece
    $\tilde {\cal M}_i$  is bounded by two hypersurfaces $\partial \Sigma_{i\pm}$  (colored curves) with common boundary $\gamma_i$ (red dots). The two surfaces $\partial \Sigma_{i\pm}$  are not allowed to intersect away from $\gamma_i$. The gluing is then done in a way that identifies all $\gamma_i$ and which cyclicly identifies $\Sigma_{i+}$ with $\Sigma_{(i+1)-}$.  In the figure, we have given  $\Sigma_{i+}$ and $\Sigma_{(i+1)-}$ the same color to make the cut-and-paste visually clear.
    }
\end{figure}

We wish to cut out pieces from each of the ${\cal M}_i$, and we would like each piece to reach $\gamma_i$.  To do so, in each ${\cal M}_i$ we choose two smooth hypersurfaces with boundary.   These hypersurfaces will be called $\Sigma_{i\pm}$. We require the boundary $\partial \Sigma_{i\pm}$ of each hypersurfaces to be precisely $\gamma_i$, but we forbid $\partial \Sigma_{i+}$ from intersecting $\partial \Sigma_{i-}$ in their interiors; see left panel of figure \ref{fig:cutandpaste}.  We suppose that the (singular) surface $\Sigma_{i} = \Sigma_{i+} \cup \Sigma_{i-}$ divides  ${\cal M}_i$ into two pieces and pick one of them to call $\tilde {\cal  M}_i$.  We will have no further need of the other piece.

At this point, we require that the induced geometry on $\Sigma_{i+}$ agrees (cyclically) with that on $\Sigma_{(i+1)-}$.  In particular,  we require there to be diffeomorphisms $\chi_i: \Sigma_{i+} \rightarrow \Sigma_{(i+1)-}$ that preserve the induced metric and where the restriction of $\chi_i$ to $\gamma_i$ is precisely $\psi_i \circ \psi^{-1}_{i+1}$.  We may then use the maps $\chi_i$ to paste the pieces $\tilde {\cal  M}_i$ together by identifying each $\Sigma_{i+}$ with the succeeding $\Sigma_{(i+1)-}$; see again figure \ref{fig:cutandpaste}.

The resulting spacetime $\tilde {\cal M} = \cup_i \tilde {\cal M}_i$ is may be singular at the surfaces $\Sigma_{i\pm}$. Away from the
common image $\tilde \gamma$ of the $\gamma_i$, the issue is just a potential discontinuity in the extrinsic curvature across each $\Sigma_{i\pm}$. This would lead to a delta-function in the Riemann tensor, with the delta-function supported on $\Sigma_{i\pm}$.  The situation at $\tilde \gamma$ is more subtle, but the above conditions nevertheless imply that the metric on $\tilde {\cal M}$ remains continuous at $\tilde \gamma$.  In particular, for any codimension there is a well-defined notion of surfaces in $\tilde {\cal M}$ that intersect $\tilde \gamma$ orthogonally. Below, we will sometimes refer to $\tilde \gamma$ below as the {\it splitting surface} for $\tilde {\cal M}$.

If we were to smooth out the singularity of $\tilde {\cal M}$ at the splitting surface $\tilde \gamma$ (perhaps using a complex metric if necessary), then in the limit where the smoothing is removed we could arrange for the curvatures to become large only in the planes $\Sigma^\perp_p$ orthogonal to $\tilde \gamma$ at each point $p \in \tilde \gamma$. As just noted, these $\Sigma^\perp_p$ are well-defined despite the singularity at $\tilde \gamma$.   We may thus follow \cite{Louko:1995jw} and \cite{Neiman:2013ap} in using the two-dimensional Gauss-Bonnet theorem to compute what will amount to delta-functions at $\tilde \gamma$ for the associated components of the Riemann tensor.

Now, the Einstein-Hilbert action is the integral of $\sqrt{-g} R$.
As explained in \cite{Colin-Ellerin:2020mva}, the end result is that $\sqrt{-g} R$ will contain a Dirac delta-function on the codimension-2 surface $\tilde \gamma$. To write this in a convenient form, let us choose a one-parameter family of neighborhoods $\mathcal{U}_\epsilon \supset \tilde \gamma$ with topology $D \times \tilde \gamma$ where $D$ is a disk.  If the $\mathcal{U}_\epsilon$ shrink to $\tilde \gamma$ as $\epsilon \rightarrow 0$, we may use the above-mentioned form of the Gauss-Bonnet theorem to write the Einstein-Hilbert action in the form\footnote{Ref \cite{Colin-Ellerin:2020mva} considered a Schwinger-Keldysh contour for which the path integral involves $\exp \left( i \int \eta \sqrt{-g} R\right)$ where $\eta = \pm 1$ changed sign at the splitting surface $\tilde \gamma$ due to $\tilde \gamma$ lying on a time-fold.  Here we instead assume that the sign $\eta$ is always $+1$ near the splitting surface.  The case $\eta = -1$ is just the complex conjugate.  In fact, by introducing appropriate time-folds, one can arrange for any path integral to have any of these 3 behaviors for $\eta$ in the region near $\tilde \gamma$.}
\begin{eqnarray}
\label{eq:EH}
S_{EH} &=& \frac{1}{16\pi G_N}\int_{\tilde {\cal M}} \sqrt{-g} R \\ &:=& \lim_{\epsilon \rightarrow 0} \left(
\frac{1}{16\pi G_N} \int_{\tilde {\cal M}\setminus \mathcal{U}_\epsilon} \sqrt{-g} R -  \frac{1}{8\pi G_N} {\cal P} \int_{\partial \mathcal{U}_\epsilon}  \ \sqrt{|h|} K  \right) + i(\frac{{\cal N}}{4}-1) \frac{A[\tilde \gamma]}{4 G_N}.
\end{eqnarray}

Again, the argument is that the only important contributions from $\mathcal{U}_\epsilon$ involve the planes $\Sigma^\perp_p$ orthogonal to $\tilde \gamma$, and that those contributions can be computed using the Gauss-Bonnet theorem.  However, integrating over $\tilde \gamma$ allows the result to be written in the form \eqref{eq:EH} which displays the higher dimensional covariance.
Thus in \eqref{eq:EH} the symbol $K$ denotes the usual codimension-1 traced extrinsic curvature of $\partial \mathcal{U}_\epsilon$.  This quantity will diverge when $\partial \mathcal{U}_\epsilon$ becomes null and, indeed, the integrand $\sqrt{|h|} K$ in the second term will typically have a pole at such points when expressed as an integral over a coordinate $\lambda$ that agrees locally with the affine parameter along the null tangent to  $\partial \mathcal{U}_\epsilon$.  The symbol ${\cal P}$ indicates that we should take the principle part of the resulting integral over $\lambda$, so that the positive and negative divergences cancel to leave a result that is finite (and real).

\begin{figure}[h!]
    \centering
    \vspace{1cm}
    \includegraphics[width=0.25\textwidth]{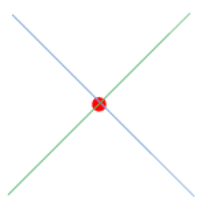}
    \caption{\label{fig:cong} In a smooth Lorentz-signature spacetime, every codimension-2 surface (red dot) is approached by four orthogonal null congruences.  These approach from future-left and past right (both blue), and from future-right and past left (both green).
    }
\end{figure}

However,  the form of the complex Gauss-Bonnet theorem  stated in \cite{Colin-Ellerin:2020mva} requires using an $i\epsilon$ prescription for $\sqrt{|h|} K$ that gives an additional non-zero imaginary contribution not present in the principal part discussed above.  Such imaginary parts can be computed explicitly and yield the net result $i\frac{{\cal N}}{4} \frac{A[\tilde \gamma]}{4 G_N}$.  In \eqref{eq:EH}, we have
combined this with the contribution $-i \frac{A[\tilde \gamma]}{4G_N}$ to the Gauss-Bonnet theorem that comes from fact that a disk has Euler character $1$ to obtain the final term in \eqref{eq:EH}.
simply written as a separate term in \eqref{eq:EH}.  Here ${\cal N}$ is the number of non-intersecting null congruences that approach $\tilde \gamma$ orthogonally.  For example, any smooth Lorentz-signature spacetime has ${\cal N}=4$ for any $\tilde \gamma$, corresponding to null congruences that approach from future-right, future-left, past-right, and past-left as shown in figure \ref{fig:cong}.  Thus in that case, when combined with the term $-i \frac{A[\tilde \gamma]}{4G_N}$ that comes from fact that a disk has Euler character $1$, we find $\frac{\cal N}{4}-1 =0$ and the action is real.  But in general our cut-and-paste construction can give ${\cal N} \neq 4$ at a general splitting surface $\tilde \gamma$ so that, even for real Lorentz-signature metrics, there can be a net imaginary contribution to \eqref{eq:EH} .

As described in \cite{Louko:1995jw,Neiman:2013ap,Colin-Ellerin:2020mva}, the above definition is a natural one in many ways.  However, since the metric on $\tilde {\cal M}$ is real, the sign of the final (imaginary) term is the result of an external input.  It is made so as to suppress contributions in the path integral from those $\tilde \gamma$ with large ${\cal N}$, and thus to agree with results of standard Euclidean computations; see e.g. \cite{Colin-Ellerin:2020mva,Colin-Ellerin:2021jev}.  We simply take this sign as a definition of our Lorentz-signature action and proceed.

The above construction is easy to generalize to allow any number $n_S$ of non-intersecting splitting surfaces. We assume this generalization below. It would be natural to also allow splitting surfaces to intersect, and indeed to form branched networks.  But we leave the study of such intersections for future investigation.

\section{Saddle-point methods in general dimension}
\label{sec:thimbles}

Having defined the space of paths over which we will integrate,  we will shortly wish to analyze our Lorentz-signature path integral in the semiclassical approximation.  Before doing so, however, it is useful to first take a moment to note certain results from the literature on saddle-point methods.  We also make a small observation that will prove useful in our main analysis in section \ref{sec:main} below.

The saddle-point approximation is familiar to every student of modern physics.  And for a single variable of integration, it is also familiar that the mere existence of a saddle-point $p$ does not guarantee its relevance to the semiclassical limit of the given integral.  Indeed, to be relevant it must be possible to deform the original contour of integration, without passing through singularities, so that the deformed contour follows the steepest descent contour through $p$,  at least to a suitable extent.

Checking this condition can be complicated, and generally becomes even more so for higher dimensional integrals (involving several variables of integration).  Here we will be interested in the infinite-dimensional limit that defines our path integral.  Luckily, however, there are theorems from either Morse theory or Picard-Lefshetz theory that greatly simplify the analysis for the particular case to be studied below.

The relevant results are explained in \cite{FAs,FP,AGV,BH,BH2,H} and summarized in \cite{Witten:2010cx}, whose presentation we will largely follow and to which we refer the reader for details. Here we suppose that we are interested in an integral over some list of integration variables $x_i$ for $i=1,\dots, d$, and that each is integrated over the real line\footnote{The reader may ask if the path integral over Lorentz-signature metrics is in fact of this form, as the constraint on the signature means that the space of allowed metrics has a finite boundary.  One may avoid this issue by thinking of the path integral as integrating over vielbein fields, which are essentially a square-root of the metric taken so as to ensure Lorentz signature always.  But we will assume here that the semiclassical limit is not sensitive to such subtleties.}.  Thus we integrate over the real contour $\Gamma_R$ in the corresponding $n$-dimensional complex plane $\mathbb{C}^n$ and we regard our path integral as a formal $n\rightarrow \infty$ limit.

The first important fact is that every stationary point $p$ of the action is associated with two other contours of interest called ${\cal J}_p$ and ${\cal K}_p$, both of which again have the same real dimension as $\Gamma$.  They are defined so that the phase of our integrand is constant along both contours.  The first contour, ${\cal J}_p$, is the descent contour which contains all points that can be obtained by using the magnitude of the integrand to generate a gradient flow and following this flow downward from $p$.  Similarly, the second
contour ${\cal K}_p$ is the ascent contour which contains all points that can be obtained by using the magnitude of the integrand to generate a gradient flow and following this flow upward from $p$.  The relevant theorem then states that, without changing the value of the integral, $\Gamma$ can be deformed to a contour $\tilde \Gamma$ consisting of $n_p$ copies of each ${\cal J}_p$, where $n_p$ is the intersection number of ${\cal K}_p$ and $\Gamma$.  Thus a given saddle $p$ contributes precisely when $n_p \neq 0$.

We will then make use of the following further observation.  Suppose that the original integral is oscillatory, in the sense that the magnitude of the integrand is constant along $\Gamma$.  Suppose also that the saddle $p$ happens to lie on the original contour $\Gamma$.   Then the ascent contour ${\cal K}_p$ clearly intersects $\Gamma$, and must do so transversely since the magnitude of the integrand is constant along $\Gamma$ (and thus does not ascend along $\Gamma$).  This then contributes a local intersection number $\pm 1$ to $n_p$.

Furthermore, the fact that ${\cal K}_p$  ascends from $p$ means that the magnitude of the integrand at all points $q\neq p$ on ${\cal K}_p$ is strictly greater than at $p$, and thus also greater than the (constant) magnitude on $\Gamma$. As a result,  ${\cal K}_p$ can have no other intersections with $\Gamma$.  Thus the only contribution to $n_p$ comes from $p$ itself, and we find $n_p = \pm 1$.  In particular, in this context we have established that the saddle at $p$ makes a non-zero contribution in the semiclassical limit.

This is the key observation to be used in section \ref{sec:main} below.  In particular, for a fixed number $n_S$ of conical singularities, the action of any spacetime can be computed using \eqref{eq:EH} and adding appropriate (and necessarily real) matter terms.  All terms in \eqref{eq:EH} are manifestly real except for the final term, so it is only this final term that controls the magnitude of our integrand and we find
\begin{equation}
\label{eq:magnitude}
|e^{iS}| = \sum_{i=1}^{n_S} e^{\left( 1 - \frac{{\cal N}_i}{4}\right) \frac{A[\gamma_i]}{4G}},
\end{equation}
where we have allowed for $n_S$ distinct conical singularities $\gamma_i$.
Note that \eqref{eq:magnitude} involves the intrinsically discrete parameters ${\cal N}_i$, which label the number of null congruences that approach the conical singularity $\gamma_i$.  Thus the contour of integration in fact consists of an infinite number of distinct contours, one for each choice of $n_S$ and the ${\cal N}_i$ for $i=1,\dots, n_S$.   Furthermore, for given $n_S$ and ${\cal N}_i$, the magnitude \eqref{eq:magnitude} depends only on the areas $A[\gamma_i]$.  As a result, if we define sub-contours $\Gamma_{A_1, \dots , A_{n_S}}$ with fixed values of $n_S$ and the corresponding $A[\gamma_i]$, then the magnitude of our integrand is necessarily constant along each $\Gamma_{A_1, \dots , A_{n_S}}$.  Thus, as just argued above, any saddle lying on our real contour will, for appropriate $n_S$, $A[\gamma_i]$,  contribute to the semi-classical approximation of the integral over $\Gamma_{A_1, \dots , A_{n_S}}$ with weight $n_S = \pm 1$.  It will then remain only to analyze the remaining finite-dimensional set of integrals over the $A[\gamma_i]$ and any other finite set of continuous parameters that we may choose to fix below, as well as to perform the sum over $n_S$.

\section{The canonical partition function as a Lorentz-Signature Path Integral}
\label{sec:main}

As described in the introduction, the partition function of a stable non-gravitational system may be represented as the Lorentz-signature path integral \eqref{eq:EfromL} by using an appropriate weighting function $f_\beta(T)$.  We will now investigate the analogous construction \eqref{eq:betaT} in gravitational systems and evaluate the result in the semiclassical limit.  In doing so we will see that, if the Lorentzian path integral is restricted to an integral over a codimension-2 subcontour, the semiclassical approximation is controlled by a singular generalization of the standard Euclidean black hole saddles.  Using this result to perform the final two integrals then indicates the standard smooth Euclidean black hole saddles contribute with non-zero weight to the final partition function when the corresponding black holes have positive specific heat.  Our main argument is presented in section  \ref{sec:submain} below, after which sections \ref{sec:rot}-\ref{sec:hd} provide additional comments on the grand canonical ensemble for rotating and charged black holes, the microcanonical ensemble, extreme black holes, and higher derivative corrections.

\subsection{Main Argument}
\label{sec:submain}

For convenience, we restrict the discussion in this section to Einstein-Hilbert gravity with cosmological constant and minimally coupled matter.  However, we will comment on the extension to theories with higher derivative corrections in section \ref{sec:hd}.  We also find it useful to restrict attention to contexts with timelike boundaries, which might be either an asymptotically locally AdS boundary or a finite-distance boundary representing the walls of an idealized reflecting cavity\footnote{\label{foot:BC1}It is an interesting question whether the full non-linear theory is physically sensible even at the classical level with finite-distance Dirichlet boundary conditions; see \cite{Anderson2008,Anderson:2007jpe,Witten:2018lgb,Fournodavlos:2020wde,Fournodavlos:2021eye} for discussions of mathematical issues and  \cite{Andrade:2015qea} for discussion of more physical issues, though this setting has been the subject of much recent exploration in the context of AdS/CFT \cite{McGough:2016lol} and references thereto.}.  But it is then trivial to obtain the asymptoically flat case by setting the cosmological constant to zero and then taking a limit where a timelike cavity wall recedes to infinity.

Let us now begin by defining the $Z_L(T)$ that appear in \eqref{eq:betaT}.  We take these to be given by a one-parameter family of Lorentz-signature path integrals over spacetimes of the form described in section \ref{sec:prelim} and with boundaries that are topologically $Y \times S^1$.  For each path integral, the boundary metric on $Y \times S^1$ is also fixed, though we will take this metric to depend on $T$ in a manner specified below. We require $Y \times S^1$ to have a timelike Killing field $\xi_\partial$ with closed orbits that wrap the $S^1$ factor at each point on $Y$.  But since we have not required the metric on $Y \times S^1$ to be a metric product, this $\xi_\partial$ need not be hypersurface orthogonal.

Let us fix conventions by choosing a particular such boundary metric to define $Z_L(T)$ for $T=1$ and normalizing $\xi_\partial$ so that the associated Killing time is periodic with period $T=1$; i.e., so that the Killing parameter runs over $[0,1]$ along any orbit of the Killing field.  Note that this boundary metric can be reconstructed by considering any closed hypersurface $\Sigma$ (say, diffeomorphic to $Y$) and using the induced metric on $\Sigma$, together with the vector field $\xi_\partial$ on $\Sigma$ and the knowledge that the Killing time has period $T=1$.  We then take $Z_L(T)$ to be defined using a boundary metric with identical data on $\Sigma$ but where the Killing time has period $T$.

In the context of the AdS/CFT correspondence, such a path integral would indeed compute
${\rm Tr} \left( e^{iHT}\right)$ in the dual field theory. It is an interesting and deep issue whether one can more generally prove that gravitational path integrals with periodic boundaries do indeed represent traces over some Hilbert space; see e.g. comments in \cite{Harlow:2018tqv,Marolf:2020xie}.

We now wish to study the semiclassical limit of
\begin{equation}
\label{eq:beta}
Z(\beta) = \int dT f_\beta(T) \  Z_L(T)
\end{equation}
for appropriate $f_\beta(T)$.  As discussed in section \ref{sec:prelim}, our Lorentz-signature path integrals involve a sum over the number $n_S$ of codimension-2 conical singularities in the spacetime.  We will argue below that Euclidean black hole contributions come from the sector with $n_S=1$.  There will generally be semiclassical contributions from other sectors as well, especially from the $n_S=0$ sector.  As always, it is separate question to ascertain which saddle actually dominates the partition function.  For example, the classic Hawking-Page transition \cite{Hawking:1982dh} in asymptotically anti-de Sitter spacetimes is described by an exchange of dominance between a black hole solution (which for us is $n_S=1$) and periodically identified empty Euclidean AdS (which has $n_S=0$).

It is instructive to first discuss the rather trivial way in which periodic Euclidean AdS emerges as an $n_S=0$ saddle for \eqref{eq:beta} with appropriate choices of the bulk action and the boundary manifold $Y\times S^1$.  In doing so, we will focus on showing that this saddle contributes to the semiclassical approximation with non-zero weight.    We begin by first using the semiclassical approximation to evaluate the integrals that define $Z_L(T)$.  This means that we seek solutions to the classical equations of motion which have Lorentz-signature $Y\times S^1$ boundaries with period $T$. Let us suppose that the dynamics allows a stationary empty AdS solution with boundary $Y \times {\mathbb R}$, such that translations along the stationary Killing field act on the boundary by shifting ${\mathbb R}$ while leaving points on $Y$ fixed.  Then we may clearly compactify this solution to match our $Y \times S^1$ boundary conditions with any period $T$.  Furthermore, these solutions are described by real Lorentz-signature metrics and so lie on the original contour of integration.     Finally, since the Lorentzian action is real for $n_S=0$, the integrand of our path integral has constant magnitude $|e^{iS} |=1$ on the $n_S=0$ contour.  It thus follows from the observation at the end  of section \ref{sec:thimbles} that any such saddle $p$ contributes to the semiclassical approximation for $Z_L(T)$ with non-zero weight $n_p = \pm 1$.  Other periodic Lorentz-signature solutions (see e.g. \cite{Horowitz:2014hja}) will also contribute, though we will not explore such effects here\footnote{The analytic continuation of such solutions to Euclidean signature boundary conditions should also contribute to standard Euclidean path integral analyses.  It is interesting that this does not appear to have been previously studied in the literature.   But if there is an appropriate positive action theorem, the associated saddles will in any case always be subleading compared with empty AdS space.}.

Since empty AdS is stationary, the action is proportional to $T$.  Calling the coefficient $-E$, our saddle contributes $e^{-iET}$ to $Z_L(T)$.  Inserting this into \eqref{eq:beta} and integrating over $T$ gives the expected $\tilde f_{\beta}(E) := e^{-\beta E}$.
While we defined $f_\beta(T)$ to allow us to perform the $T$ integral without computation, it is instructive to examine the details.
In particular, let us choose $f_\beta(T) = \frac{1}{2\pi i}\frac{e^{E_0(-\beta +iT)}}{T+i\beta}$.  Then
\begin{equation}
\label{eq:pole}
\int dT f_\beta(T) e^{-i E T} = \frac{1}{2\pi i} \int dT \frac{e^{-\beta E_0}e^{-i(E-E_0)T}}{T+i\beta}.
\end{equation}
But for $E > E_0$ we may close the given (real) contour in the lower half of the complex $T$ plane so that Cauchy's theorem reduces evaluation of the integral to computing the residue at $T=-i\beta$.  As desired, the result is $e^{-\beta E}$, but we also see that this comes entirely from the region near $T=-i\beta$.  We may thus think of it as arising from the classical Euclidean solution given by analytically continuing our periodic empty AdS to $T=-i\beta$.

We would now like to give an analogous argument using black holes in the sector $n_S=1$, where each spacetime contains a non-trivial codimension-2 conical singularity.  However, this sector {\it cannot} contain saddle points for $Z_L(T)$.  This is because, as discussed in section \ref{sec:prelim}, the presence of a non-trivial conical singularity requires a delta-function in the Ricci scalar.  But in Einstein-Hilbert gravity coupled to familiar matter fields such delta-functions are forbidden by the equations of motion.  And the same will remain true when higher derivative corrections are included.

On the other hand, it turns out that we can find configurations that are saddles for {\it most} of the integrals that define $Z_L(T)$.  In particular, one of the integrals that defines $Z_L(T)$ can be taken to be an integral over the area $A[\tilde \gamma]$ of the conical singularity.  Let us first fix some arbitrary value of $A[\tilde \gamma]$ and perform the remaining integrals that define $Z_L(T)$, after which we will later return to integrate over $A[\tilde \gamma]$. Since we have not yet integrated over $A[\tilde \gamma]$, saddles for these integrals need not satisfy one of the Einstein equations at $\tilde \gamma$.  In particular, based on analogous Euclidean-signature analyses in e.g. \cite{Carlip:1993sa} and \cite{Dong:2019piw}, we expect that this allows the freedom to include an arbitrary conical singularity at $\tilde \gamma$ of the form discussed in section \ref{sec:prelim}.  For a given value of $A[\tilde \gamma]$, our task should then simply be to adjust the strength of this conical singularity so that there is a saddle in which $\tilde \gamma$ has the desired area.  We will refer to these spacetimes as fixed-area saddles below to distinguish them from geometries that satisfy the full Einstein equations everywhere.  Rather than attempt to rigorously characterize general Lorentz-signature fixed-area saddles in detail, we will simply proceed to first construct candidate such saddles and to then show that they are indeed stationary points of \eqref{eq:EH} under first-order variations that preserve $A[\tilde \gamma]$.

The interesting observation is that fixed-area saddles of this sort do in fact generally exist on what we call the original  real Lorentz-signature $n_S=1$ contour.  In particular, let us begin by considering any stationary black hole exterior with a Killing horizon for which the horizon-generating Killing field $\xi$ both preserves a boundary of the form $Y \times {\mathbb R}$ and agrees there with $\xi_\partial$. This condition will in particular constrain the angular velocity of the black hole when $Y$ admits rotational symmetries.  And when coupled to a Maxwell field we would also fix the electric or magnetic potential on the boundary. However, the boundary metric does not constraint the Killing energy $E_\xi$, so we can generally adjust this parameter to obtain an exterior geometry ${\cal M}_{A}$ in which the horizon area $A$ agrees with area $A[\tilde \gamma]$ we wish to fix at our conical singularity $\tilde \gamma$.  In the main discussion below, we will assume ${\cal M}_{A}$  to have a bifurcate Killing horizon (so that the surface gravity $\kappa$ is non-zero), though we will include brief comments on the extreme $\kappa=0$ case in section \ref{sec:extreme}.  The singularity $\tilde \gamma$ will appear shortly in the next step of our construction.

\begin{figure}[h!]
    \centering
    \vspace{1cm}
    \includegraphics[width=0.25\textwidth]{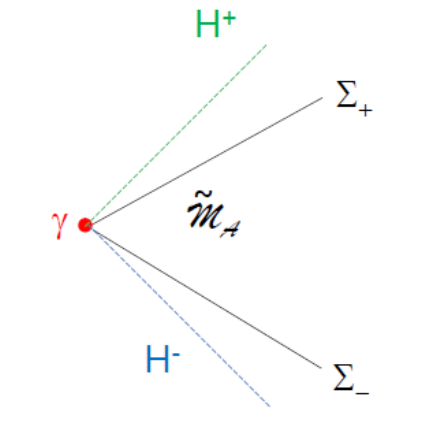}
    \caption{\label{fig:ext} We take our black hole exterior ${\cal M}_A$ to include the bifurcation surface $\gamma$ (red dot), but not the past horizon $H^-$ (dashed blue) or the future horizon $H^+$ (dashed green).  As a result, the quotient by $e^{\xi T}$ may also be described by introducing two slices $\Sigma_{\pm}$ (each at a constant Killing time $\pm T/2$), focussing on the region $\tilde {\cal M}_A$ between them, and identifying $\Sigma_+$ with $\Sigma_i$.  This is a special case of the cut-and-paste construction of section \ref{sec:prelim} using only a single spacetime ${\cal M}_1 = {\cal M}_A$.  Note that any geodesic that remains in the region $\tilde {\cal M}_A$ and approaches the bifurcation surface must do so in a spacelike manner. As a result, the quotient ${\cal M}_{A,T}$ contains no null congruences that approach the image of the bifurcation surface.  This means that ${\cal M}_{A,T}$  has ${\cal N}=0$ in the notation of section \ref{sec:prelim}.  Recall also that the image of $\gamma$ in ${\cal M}_{A,T}$  will be called $\tilde \gamma$.
    }
\end{figure}

We take the exterior ${\cal M}_{A}$ include the bifurcation surface, but not the future or past horizons; see figure \ref{fig:ext}. Taking a quotient of this exterior by the diffeomorphism\footnote{This is the diffeomorphism that moves every point along its Killing orbit by a Killing parameter $T$.} $e^{\xi T}$ then yields a spacetime  ${\cal M}_{A,T}$ on which $\xi$ continues to generate an isometry.  However, the new isometry on ${\cal M}_{A,T}$  has $U(1)$ orbits that contract to points at the image of the bifurcation surface for the original black hole exterior ${\cal M}_A$.  We will henceforth denote this image as $\tilde \gamma$ and, for convenience, we may sometimes continue to refer to it as a bifurcation surface for ${\cal M}_{A,T}$ even though $\tilde \gamma$ has no orthogonal null congruences in ${\cal M}_{A,T}$; see again figure \ref{fig:ext}.

As also illustrated in figure \ref{fig:ext}, the quotient ${\cal M}_{A,T}$ lies in the class of spacetimes described in section \ref{sec:prelim}, and thus it lies on the original real Lorentz-signature contour of integration for $n_S=1$.
So, if it does indeed define a fixed-area saddle, it will necessarily contribute to the semiclassical evaluation of our fixed-area path integral as described at the end of section \ref{sec:thimbles}.

% Furthermore, the action of any spacetime on this contour can be computed using \eqref{eq:EH}, which involves the intrinsically discrete parameter ${\cal N}$.  Thus our contour in fact consists of an infinite number of distinct contours, one for each

 %Furthermore, it is clear from figure XXX\DM{Which fig?} that $\gamma$ can be only approached from ${\cal M}_{A,T}$ along curves that are spacelike at $\gamma$. In particular, there are no null congruences that approach $\gamma$ and ${\cal N}=0$ in \eqref{eq:EH}. In particular, ${\cal N}=0$

%since we have fixed the area $A[\gamma]$ of the conical singularity,

Now, by construction, ${\cal M}_{A,T}$ solves the equations of motion away from $\tilde \gamma$.  Furthermore, most of the equations of motion will also hold at $\tilde \gamma$ by continuity.  The one subtlety in this argument is that, as described in section \ref{sec:prelim}, the Riemann tensor of ${\cal M}_{A,T}$ turns out to be the sum of two terms.  One term is identical to the Riemann tensor of the parent space ${\cal M}$, and this term is indeed continuous.  The other term is a delta-function of some constant (complex) amplitude supported on $\tilde \gamma$.

To some readers it will now be readily apparent that ${\cal M}_{A,T}$ is indeed a fixed-area saddle.  This may be especially clear by analogy with the Euclidean discussion in e.g. \cite{Carlip:1993sa}.  However, for those who are interested we provide the details of the Lorentz-signature argument for ${\cal M}_{A,T}$ in appendix \ref{sec:FAS}.
This then establishes that ${\cal M}_{A,T}$ contributes with non-zero weight to the semiclassical evaluation of our fixed-area path integral.

As usual, and as forewarned above, the question of whether this contribution in fact dominates the fixed-area path integral will generally require further investigation.  In particular, in many cases there will be several stationary black hole exteriors with boundary $Y \times {\mathbb R}$ and having the same $A$.  They then give rise to a corresponding number of quotients ${\cal M}_{A,T}$, of which at most one can domiante. This multiplicity is often associated with the topology of the black hole horizon. Examples of this sort arise in AdS$_d \ \times \ X$ for compact $X$.  For appropriate $Y \times S^1$, one saddle will dimensionally reduce to the usual AdS$_d$-Schwarzschild black hole, while other saddles will localize to various extents within the $X$ factor.   We will not explore the associated phase structures here, though our arguments below will ensure that the implications for the full partition function are the same as in standard Euclidean analyses.

Note, however, that the construction above does not generally yield families of solutions with additional continuous parameters.  In particular, when $Y$ is a sphere, one might ask what happens if one attempts to hold $A$ constant while changing the angular momentum $J$ of the black hole.  The answer is that this changes the angular velocity $\Omega$ of the black hole as well, so that the horizon generating Killing field is now some $\chi = \partial_t + \Omega' \partial_\phi$ with $\Omega' \neq \Omega$.  But since we have not changed the metric on the boundary $Y \times S^1$, the Killing field along the boundary $S^1$ factor remains $\xi_\partial = \partial_t + \Omega \partial \phi$ with the original value of $\Omega$.  So the analogue of the above quotient construction would still continue to identify the new black hole exterior under translations generated by $\xi = \partial_t + \Omega \partial_\phi$ in the bulk, which differs from $\chi$ by $(\Omega - \Omega')\partial_\phi$.  As a result, $\xi$ will now map the bifurcation surface to itself in a non-trivial way, and the quotient will no longer be of the form described in section \ref{sec:prelim} above\footnote{The astute reader may note that, since we have not required it to act orthogonally to the $Y$ factor in any sense, the boundary Killing field $\xi_\partial$ that acts along the $S^1$ factor is generally not unique.  For example, on $S^3 \times S^1$ one can add to any $\xi_\partial$ any integer multiple of a $2\pi$ rotation on the $S^3$.   This leads to an additional discrete family of solutions for each $A,T$, analogous to those studied in Euclidean signature in e.g. \cite{Dijkgraaf:2000fq,Maloney:2007ud}}.

We now wish to evaluate the Lorentz-signature action on each ${\cal M}_{A,T}$, and to write the result in a useful form.  Here again we may refer to the results of section \ref{sec:prelim}, since ${\cal M}_{A,T}$ is of the form studied there.  Since ${\cal N}=0$, we may thus use \eqref{eq:EH} to write the full Lorentz-signature action $S$ of ${\cal M}_{A,T}$ in the form
\begin{eqnarray}
\label{eq:SL}
S =  &\lim_{\epsilon \rightarrow 0}& \left[
\frac{1}{16\pi G_N} \int_{{\cal M}_{A,T}\setminus \mathcal{U}_\epsilon}  \sqrt{-g} \left( R + 16 \pi G_N  L_{matter} \right) -  \frac{1}{8\pi G_N} {\cal } \int_{\partial \mathcal{U}_\epsilon}  \ \sqrt{|h|} K   \right] \\ &+& \int_{\partial {\cal M}_{A,t}} {\cal B}\  - \ i \frac{A[\tilde \gamma]}{4 G_N}.
\end{eqnarray}
Here $L_{matter}$ is the matter Lagrange density (which we take to include any cosmological constant term), and ${\cal B}$ describes whatever boundary terms are required at ${\partial {\cal M}_{A,t}}$. Note that we have dropped the principal part symbol ${\cal P}$ that in \eqref{eq:EH} acted on the Gibbons-Hawking term at ${\partial \mathcal{U}_\epsilon}$.  This is possible here since we can choose ${\partial \mathcal{U}_\epsilon}$ to be everywhere timelike so that the integrand never diverges.  Indeed, we can choose $\mathcal{U}_\epsilon$ to the be the region within some geodesic distance $\epsilon$ of $\tilde \gamma$, in which case the integrand is invariant under all symmetries of the black hole.  In particular, for a Schwarzschild-(A)dS black hole in standard coordinates, the quantity $\sqrt{|h|} K$ is constant on $\mathcal{U}_\epsilon$.

Recall now that we consider only minimally coupled matter fields so that the matter action may be written entirely in terms of fields and their  first derivatives.  Such fields are bounded near $\tilde \gamma$.
Since the volume $\int_{\partial \mathcal{U}_\epsilon}  \ \sqrt{|h|}$ of the interior boundary vanishes as $\epsilon \rightarrow 0$, this means that any corresponding matter boundary terms on ${\partial \mathcal{U}_\epsilon}$ vanish as $\epsilon \rightarrow 0$.  We are thus free to add such terms to \eqref{eq:SL} as well.   In particular, we may add the matter boundary term that promotes $\int L_{matter}$ to a good variational principle for some class of boundary conditions on ${\partial \mathcal{U}_\epsilon}$.  Doing so, and combining it with the first three terms in \eqref{eq:SL} defines an action $S({\cal M}_{A,T}\setminus \mathcal{U}_\epsilon)$ for a gravity-plus-matter system on some manifold $X \times S^1$ where $\partial X$ is homeomorphic to $Y \cup \tilde \gamma$.  In particular, the $S^1$ factor does not degenerate anywhere in this ${\cal M}_{A,T}\setminus \mathcal{U}_\epsilon$.  We can thus write this action in the Hamiltonian form
\begin{equation}
\label{eq:can}
S({\cal M}_{A,T}\setminus \mathcal{U}_\epsilon)  = \int_{S^1} \left( -H + \int_X p_\alpha \dot{q}^\alpha\right)
\end{equation}
in terms of appropriate coordinates $q^\alpha$ on the space of fields at each spacetime point and their conjugate momenta $p_\alpha$.  Furthermore, since our action describes Einstein-Hilbert gravity with minimally coupled matter, and since we have an explicit Gibbons-Hawking term at $\partial  \mathcal{U}_\epsilon$, the Hamiltonian $H$ can be written as an integral of constraints together with boundary terms at $\partial {\cal M}_{A,T}$.

Let us now recall that $H$ is the Hamiltonian that generates translations along $\xi$.  This means that the boundary term at $\partial \mathcal{U}_\epsilon$ includes a factor of $\xi$ evaluated near the bifurcation surface $\tilde \gamma$.  But since $\xi$ vanishes at $\tilde \gamma$, the limit $\epsilon \rightarrow 0$ of the gravitational Hamiltonian boundary term at $\partial \mathcal{U}_\epsilon$ will vanish\footnote{An important point here is that the component $\Pi^{\xi \xi} = \Pi^{ij}\xi_i \xi_j$ of Brown-York stress tensor $\Pi_{ij}$ at $\partial \mathcal{U}_\epsilon$ vanishes as fast as the norm $\xi^i \xi_i$  as $\epsilon \rightarrow 0$.  Since $\Pi_{ij} = \frac{1}{8\pi G}(K_{ij} - K h_{ij})$, where $h_{ij}$ is the induced metric on $\partial \mathcal{U}_\epsilon$, this is the one component of $\Pi_{ij}$ in which the extrinsic curvature component $K_{\xi \xi}$ does not appear. This $K_{\xi \xi}$ is large in the sense that $\frac{K_{\xi \xi}}{\xi^i \xi_i}$ diverges, so it represents a large extrinsic curvature in any orthonormal frame. The appearance of this component in the trace ($K$) is what prevents us from dropping the Gibbons-Hawking term in \eqref{eq:SL}.}.  We may thus drop it from our action and  think of  $H$ as simply being the sum of constraints and the usual boundary term at $S^1 \times Y$.  It's value on $\partial {\cal M}_{A,T}$ is thus just the usual conserved energy $E_\xi$ associated with the Killing field $\xi$ for the black hole exterior $\partial {\cal M}_{A}$ that was mentioned at the beginning of this section.

To complete the computation of \eqref{eq:can}, we take $\dot{q} = \pounds_\xi q$ and note that this vanishes due to the Killing symmetry of $\partial {\cal M}_{A,T}$.  Thus \eqref{eq:can} is simply $-E_{\xi} T$ and the contribution of our black hole quotient saddles to $Z_L(T)$ takes the form
\begin{equation}
\label{eq:exp}
Z_{L,BH}(T) = \int dA  \ e^{iS({\cal M}_{A,T})} = \int dA e^{A/4G} e^{-i E_\xi T},
\end{equation}
where we remind the reader that for a given family of black holes this $E_\xi$ is a function of the horizon area $A$.
%$= \int dE_\xi e^{A/4G} e^{i (E-\Omega J) T},$
%where the subscript $BH$ on the left-hand-side indicates that we can considered only the contributions to $Z_L(T)$ that come from our particular family of black holes ${\cal M}_{E_\xi}$.  In particular, since $\Omega$ is fixed  this family determines both $E$ and $J$ as functions of $E_\xi$.
To obtain the full partition function $Z_L(T)$, we would also need to sum over the possible families of black holes, and to include other contributions from sectors with $n_S=0$ or $n_S \ge 2$.

Note that for a non-gravitational system with a Bekenstein-Hawking density of states $e^{A/4G}dA$, our \eqref{eq:exp} would yield $Tr(e^{iH_\xi T})$, where $H_\xi$ is the operator with eigenvalues $E_\xi (A)$.
However, we noted already in the introduction that we did not expect $Z_L(T)$ to define a sensible function of $T$.  This result can now be seen explicitly since in most contexts the area $A$ of a black hole is not bounded\footnote{An interesting exception is when one studies black holes inside a cavity on whose walls that boundary metric has been fixed.  In such cases, the area $A$ of the black hole is typically bounded by the area of the cavity wall.}.  Thus the integral in \eqref{eq:exp} fails to converge at large $A$.

Instead, we expected to obtain sensible results only after integrating over $T$ in \eqref{eq:beta}.  Let us therefore consider
\begin{equation}
\label{eq:beta2}
Z_{L,BH}(\beta) = \int dA dT  \ f_\beta(T) e^{iS({\cal M}_{A,T})} = \int dA dT \ f_\beta(T) e^{A/4G} e^{-i E_\xi T},
\end{equation}
with the understanding the we should perform the $T$ integral {\it before} integrating over $A$.   As in our discussion of empty periodic AdS, this $T$ integral clearly computes the Fourier transform $\tilde f_\beta(E_\xi)$ which, just below \eqref{eq:EfromL2}, we defined to be $e^{-\beta E_\xi}$  for $E$ greater than some $E_0$.  Here we simply choose $E_0$ to be the ground state energy of our system, in the sense that it gives the infimum of the energy $E_\xi$ over all solutions with boundary conditions $Y\times {\cal R}$ as defined above\footnote{We restrict attention to systems where such an infimum exists.  If it does not, one expects the canonical ensemble to be ill-defined.}.  Thus $E_\xi \ge E_0$ for all black hole exteriors ${\cal M}_A$.

As in the discussion of periodic empty AdS around \eqref{eq:pole},  the result $e^{-\beta E}$ can be viewed as arising entirely from the region the complex $T$-plane near $T=-i\beta$.  We may thus also think of it as arising from the classical solution given by analytically continuing ${\cal M}_{A,T}$ to $T=-i\beta$.

When $\xi$ is hypersurface orthogonal this analytic continuation gives a spacetime that, away from $\tilde \gamma$, is locally identical to the usual Euclidean black hole of area $A$.  However, the period of Euclidean time has been enforced by hand to agree with the external parameter $\beta$, and thus need not agree with the preferred value $2\pi/\kappa_A$ determined by the surface gravity $\kappa_A$  of ${\cal M}_{A}$.  As a result, we generally find a conical singularity at the corresponding Euclidean horizon.  This should be no surprise since, as discussed earlier, it is the integral over $A[\tilde \gamma]$ that in the semiclassical approximation would impose the requirement that there should be no delta-function in the Ricci scalar at this horizon.  This integral has not yet been performed, but we will turn to it shortly.

Before doing so, however, we note that when $\xi$ fails to be hypersurface orthogonal the analytic continuation to $T=-i\beta$ instead yields metrics that are complex-valued.   This is familiar from, e.g., the naive analytic continuation of the Kerr solution (say, in co-rotating coordinates). Indeed, even the metric on the boundary $Y \times S^1$ is generally complex.   Although complex metrics may be unfamiliar to some readers, they are arguably the most natural way to study the thermodyanmics of rotating black holes \cite{Brown:1990fk,Brown:1990di}.   We will discuss this further in section \ref{sec:rot} below.

Our black hole contributions to the partition function may now be written
\begin{equation}
\label{eq:beta3}
Z_{L,BH}(\beta) = \int_0^\infty dA  \ e^{-\beta E_\xi} e^{A/4G},
\end{equation}
where, again, %since $\Omega$ is fixed by the boundary conditions,  both $E$ and $J$ are functions of
$E_\xi$ is a function of $A$ specified the particular family of black hole exteriors ${\cal M}_{A}$ used above.
This is precisely the standard form for the canonical ensemble partition function of a statistical mechanical system with density of states $e^{A/4G}dA$.  The usual analysis then tells us that, for small $G$, our \eqref{eq:beta3} may be approximated by a sum over local minima of the free energy $F_\xi = E_\xi - A/4\beta G$, and that such local minima are precisely those stationary points with positive `specific heat' $\frac{d E_{\xi}}{d\tau}$, where $\tau=\tau(A)$ is an effective temperature defined by $\tau = 4 G \frac{dE_\xi }{dA }$.   This suggests that smooth Euclidean black hole saddles (or complex generalizations thereof) with the specified angular velocity $\Omega$ and inverse temperature $\frac{1}{4G} \frac{dA}{dE_\xi} = \beta$ contribute to the semiclassical approximation for $Z(\beta)$ with non-zero weight when the corresponding Lorentz-signature black hole has positive specific heat (in the sense of having positive $dE_\xi/d\tau$).  We also see that there are no such contributions from those with negative specific heat, which instead correspond to local maxima of the free energy $F_\xi = E_\xi - A/4\beta G$.

Before concluding this section, we should again pause to note what we believe to be a small technical caveat.  What we actually did above was to show that smooth Euclidean black holes with positive specific heat contribute to the fixed-area path integral, and that they also define saddles that contribute to the integrals we performed over $A$ and $T$.  This, however, is not quite the same as proving definitively that they contribute with non-zero weight to the full partition function.  The issue is related to the fact that we may, perhaps, have missed equally important or more dominant contributions from saddles or constrained-saddles that we did not identify above and that, when the final integral over constrained saddles is extended to include such new contributions, it is possible that our positive specific heat black holes no longer define local maxima of the integrand (say, in the analogue of \eqref{eq:beta3}), but may perhaps give only saddle points).  It seems likely that this is related to the fact that the criterion found above for a black hole to contribute depends only on its specific heat while, on physical grounds, for e.g. charged and rotating black holes one expects the criterion to depend on the full Hessian of the free energy with respect to all thermodynamic potentials.  More will be said about this in section \ref{sec:disc} below.

\subsection{The Grand Canonical Ensemble for Rotating and charged black holes}
\label{sec:rot}

The above argument applies to all classes of black hole solutions, whether or not they have electric charge, angular momentum, or more general conserved charges.  However, it is worth commenting further on the charged and/or rotating contexts.  The comments below are standard, in the sense that similar remarks have appeared in many past discussions of black hole thermodynamics.  But we include them here for completeness and clarity.

Let us first address the case of rotation.
To this end, let us thus suppose that $Y \times S^1$ admits both a time-translation $\partial_t$ and a time-reflection symmetry that preserves the orbits of $\partial_t$.  We also suppose that $Y$ itself admits a Killing field $\partial_\phi$ so that we may define an angular velocity $\Omega$ by writing $\xi = \partial_t + \Omega \partial_\phi$.  In this case, the conserved charge $E_\xi$ associated with $\xi$ may be written in the form $E_\xi = E-\Omega J$, where (due to the usual sign conventions) $E,J$ are the conserved charges associated with $\partial_t, \partial_\phi$.  Here we remind the reader that it is sufficient in this discussion to discuss Killing fields of the boundary $Y \times S^1$ whether or not they can be extended to bulk Killing fields for any particular solution.

Similarly, if the boundary conditions on $Y \times S^1$ fix a non-zero electric potential $\Phi$, then the charge $E_\xi$ defined by writing the action in canonical form contains an explicit term $-Q\Phi$, where $Q$ is the total electric charge.  In this context it is thus natural to write $E_\xi = E - \Omega J - Q \Phi$ so that our final result \eqref{eq:beta3} becomes
\begin{equation}
\label{eq:beta4}
Z_{BH}(\beta, \Omega, \Phi) = \int_0^\infty dA  \ e^{-\beta (E - \Omega J - Q\Phi)} e^{A/4G},
\end{equation}
which takes the form of a standard grand canonical ensemble with fixed potentials $\Omega, \Phi$.

Now, as described above,  each $A$ in the integral \eqref{eq:beta4} can be associated with the analytic continuation of some ${\cal M}_{A,T}$ to imaginary Killing times. Recall that the analytically-continued metric is real when $\xi$ is hypersurface orhtogonal, but more generally it is complex.  Similarly, for nonzero electric potential $\Phi$ the analytically-continued electromagnetic vector potential will have imaginary components.

However, we can also typically also describe rotating or charged black holes using real Euclidean metrics by making additional analytic continuations of $\Omega$ and $\Phi$ as in \cite{Gibbons:1979xm,HawkEin}.   For this discussion we further assume the exterior solutions ${\cal M}_{A,T}$ to be invariant under $(t, \phi) \rightarrow (-t, -\phi)$ as is the case for the (A)dS-Kerr-Newman family of solutions and its kin.  Here it is useful to in fact consider a different analytic continuation defined by keeping $t$ real but writing $\Omega = - i \Omega_E$  and $\Phi = - i \Phi_E$ and taking $\Omega_E, \Phi_E$ real.  This yields a complex metric that is invariant under the combined action of $t \rightarrow - t$ and complex conjugation.

Since we now have $\xi = \partial_t + i \Omega_E \partial_\phi$, we may also describe this symmetry as invariance under simultaneously taking complex conjugates and changing the sign of the Killing parameter $\eta$ along the orbits of $\xi$.  As a result, the further analytic continuation $\eta \rightarrow -i \eta_E$ for real $\eta_E$ will yield a solution that is invariant under complex conjugation alone.    Thus this metric is real.  It is also manifestly Euclidean to leading order near the bifurcation surface (where the Lorentz-signature metric takes the universal Rindler-like form \eqref{eq:asymptg}).  As a result, the metric will be real and Euclidean everywhere so long as it is smooth and invertible at every point.  This is again the case, for example, for the familiar rotating and charged black holes in asymptotically flat or asymptotically AdS spacetimes.  Furthermore, since the $T$ used above is in fact the period of the Killing parameter $\eta$, this real metric is just the continuation we need to set $T=-i\beta$ with real $\Omega_E$.  To summarize then, while for real $\Omega$ our $T=-i\beta$ spacetimes are generally complex, analytically continuing the angular velocity to real $\Omega_E$ yields what are often called the (real) rotating black hole solutions in Euclidean signature\footnote{It is interesting to ask if, instead of performing the analytic continuation $\Omega_E=-i\Omega$, one could instead obtain similar results by integrating $Z(\beta, \Omega)$ against an appropriate function of $\Omega, \Omega_E$ in analogy with \eqref{eq:EfromL}. However, it is not immediately clear how to arrange the desired construction so that all integrals converge. }.

\subsection{The Microcanonical partition function}
\label{sec:micro}

So far we have studied the canonical ensemble and its rotating and charged cousins.  But the main argument above is just as easily applied to the microcanonical partition function.  This is itself an interesting statement, as when one attempts to formulate the ensembles directly in terms of Euclidean gravitational path integrals, the study of stability for the associated microcanonical saddles  in asymptotically AdS spacetimes turns out to be more subtle than for the canonical partition function \cite{Marolf:2022jra} (though they are more comparable when studied inside a reflecting cavity of finite size \cite{Marolf:2022jra,Marolf:2022ntb}).

In particular, the microcanonical ensemble can be described by choosing the weighting function $\tilde f(H)$ in \eqref{eq:EfromL2} to be a Gaussian centered at some energy $E_0$.  In the limit of vanishing width $\sigma_E$, this gives a delta-function so that ${\rm Tr} \ \tilde f(H)$ describes a microcanonical partition function.  Thus we choose
\begin{equation}
f_{E_0}(T) = e^{iTE_0} \frac{e^{-T^2/2\sigma^2}}{{2\pi}},
\end{equation}
where $\sigma = \sigma_E^{-1}$ and  where we have replaced the previous subscript $\beta$ with the $E_0$ that is more appropriate here.

Since we define it in terms of the same $Z_L(T)$ used above,
the first stages of the computation of the microcanonical partition function
\begin{equation}
Z(E_0) := \int dT f_{E_0}(T) \  Z_L(T),
\end{equation}
proceed precisely as in the canonical case to give \eqref{eq:beta2}.  Again, the $T$ integral simply computes the Fourier transform of $f_{E_0}$.  For our Gaussian, the exact result is given by stationary phase methods by solving
\begin{equation}
\label{eq:TE}
-T/\sigma^2 + iE_0 -iE_\xi = 0,
\end{equation}
or $T = -i(E_\xi-E_0) \sigma^2$.  Using this leaves us with
\begin{equation}
\label{eq:microresult}
\int_{0}^\infty  dA  e^{A/4G} \frac{\sigma}{\sqrt{2\pi}} e^{-(E_\xi-E_0)^2\sigma^2/2},
\end{equation}
where, as usual, $E_\xi$ should be viewed as a function of $A$ determined by the particular family ${\cal M}_{E_\xi}$ of black hole solutions.   As $\sigma \rightarrow \infty$, this final integral is dominated by a saddle at $E_\xi= E_0 + \epsilon$ with $\beta(E_\xi) = \epsilon \sigma^2$.  Thus we find the integral to give $e^{A(E_0)/4G}$ as $\sigma \rightarrow \infty$, which is the expected answer in the microcanonical ensemble with $\xi$-energy $E_\xi=E_0$.

Let us also briefly comment on the $\sigma^2 \rightarrow \infty$ limit of the geometries that correspond to the saddle point values $E_\xi{}*,T^*$. From \eqref{eq:TE} we have $T^* = -i(E^*_\xi-E_0) \sigma^2$.  Furthermore, thinking of \eqref{eq:microresult} in terms of an integral over $E_\xi$,  the saddle for \eqref{eq:microresult} will satisfy
\begin{equation}
\label{eq:Asaddle}
\frac{d}{dE_\xi} \frac{A}{4G}|_{E_\xi^*} = (E_\xi^* - E_0) \sigma^2.
\end{equation}
Now, the left-hand-side of \eqref{eq:Asaddle} is precisely the quantity that thermodynamics with entropy $A/4G$ would call the inverse temperature $\beta_*$ at the saddle.   In particular, $\beta_*$ is precisely the period of imaginary time that removes the potential conical singularity at $\tilde \gamma$. Again, this is because the semiclassical evaluation of the integral over $A$ imposes the equation of motion that comes from varying $A[\tilde \gamma]$, which in particular sets to zero the coefficient of any horizon delta-function in the Ricci scalar.
Thus we find $T^*=-i\beta_*$, and as $\sigma \rightarrow \infty$ \eqref{eq:Asaddle} requires $E_\xi^* \rightarrow E_0$. In this sense our microcanonical partition function is described semiclassically by precisely the usual (smooth) Euclidean black hole geometry with energy $E_0$.

\subsection{Extreme Black Holes}
\label{sec:extreme}

The analysis of section \ref{sec:submain} considered only black holes with bifurcate Killing horizons.  The reader may then ask what one finds in the zero-temperature case $\beta=\infty$, which one expects to be described by extreme black holes.  The horizons of such black holes are well-known {\it not} to be bifurcate, but instead to have an `internal infinity' at the past end of the future horizon and at the future end of the past horizon.  In the Euclidean context, the internal infinity means that extreme black holes have a different topology than non-extreme black holes, a feature which appears to lead to a surprising thermodynamic discontinuity \cite{Hawking:1994ii},  but also to interesting discussion of how this may be cured by string theory \cite{Horowitz:1996qd}.

In this context, it is interesting to note that our formulation of the canonical partition function is well-defined only for finite $\beta$.  In particular, for $E_0 \ge 0$ the integrand of \eqref{eq:pole} simply vanishes at $\beta = \infty$ due to the divergent denominator.  Thus we naturally view zero temperature as a limit, first evaluating $Z(\beta)$ for finite $\beta$ and then taking $\beta \rightarrow \infty$.    In this way our formalism avoids the above confusions and, by construction, obtains results that are continuous at zero temperature.

Of course, one can also study extreme black holes using the microcanonical ensemble.  At least as described in section \ref{sec:micro}, this
again turns out to treat extremal black holes as a limit of the non-extreme case.  That is the case because using a Gaussian (or any other smooth function) to restrict the allowed energies to a small window yields an integral over area $A$ as in \eqref{eq:microresult}, and this integral is not sensitive to the values of the integrand on sets of measure zero.

Now, the reader may note that simply taking $f_{E_0}(T)\propto e^{-iE_0}T$ would formally insert a strict delta-function into \ref{eq:microresult}.  However, since there are no black holes below exremality, in this case the delta-functions would have support only on the boundary of the region of integration.  In such cases the action of the delta-function is not well-defined, though again one may take it to be defined by a limit taken from inside the region of integration.  Doing so again renders the extreme partition function identical to the limit obtained from the non-extreme case.

\subsection{Remarks on Higher Derivative Corrections}
\label{sec:hd}

The argument above relied on using the Einstein-Hilbert action in the form \eqref{eq:EH} to evaluate contributions to the path integral from geometries with (Lorentzian) conical singularities.  Such singularities may be said to contribute delta-functions to the Riemann tensor.  In this sense they are regions of strong curvature, and the reader may ask whether this renders our argument highly sensitive to UV corrections such as higher-derivative terms, which might then contribute badly divergent powers of delta-functions.

It is thus interesting to note that by using technology from \cite{Dong:2019piw}, the argument of section \ref{sec:submain} above can also be applied to gravitational theories with perturbative higher derivative corrections.  Furthermore, in this context, the associated corrections to the final answers for any of our partition functions are indeed perturbatively small.

To understand this point, recall again that in Einstein-Hilbert gravity the area $A$ of the bifurcation surface is $4G$ times the entropy of the corresponding stationary black hole.  With higher derivative corrections, the notion corresponding to $\frac{A}{4G}$ is the Wald entropy \cite{Wald:1993nt}, which we also call the geometric entropy\footnote{Since we consider stationary spacetimes, the entropies of \cite{Dong:2013qoa} agree precisely with the Wald entropy.} $\sigma$.  It is thus natural to proceed as above replacing $\frac{A}{4G}$ by $\sigma$.

We thus consider stationary black hole exteriors ${\cal M}_{\sigma}$ and their quotients ${\cal M}_{\sigma,T}$.  This is precisely the same sort of quotient discussed above, so we obtain precisely the same sort of Lorentzian conical singularities at the would-be bifurcation surface $\tilde \gamma$.  We then wish to argue that these define stationary points of an appropriate gravitational action in which $\sigma$ has been fixed as a boundary condition.

Now, \cite{Dong:2019piw} studied fixed-$\sigma$ variational principles for Euclidean-signature metrics and argued that their stationary points are locally identical to stationary points of original higher-derivative gravity theory, except that there can be an arbitrary conical singularity on the surface\footnote{This surface is also required to satisfy an analogue of an extremal surface condition, but that condition is trivially satisfied by the singular surfaces in ${\cal M}_{\sigma,T}$ due to the Killing symmetry along $\xi$.} $\tilde \gamma$ that defines $\sigma$.  In other words, the analytic continuations of our ${\cal M}_{\sigma,T}$ are indeed stationary points.  Furthermore, \cite{Dong:2019piw} showed that the net contribution of the conical singularity to the associated {\it Euclidean} action is precisely $\sigma (m-1)$, where $2\pi m$ is the total angle subtended by a circle around the conical singularity (i.e., where $2\pi(m-1)$ is the familiar conical deficit angle). For later use we note that, even though it is associated with the singularity, the contribution $\sigma m$ to the action is proportional to the period of Euclidean time (since this is linear in $m$), but that the remaining $-\sigma$ is not. All other terms in the action describing the region away from the conical singularity are manifestly proportional to the period $\beta$ of Euclidean time.

As noted in \cite{Colin-Ellerin:2020mva}, one may formally analytically-continue the imaginary time results of \cite{Dong:2019piw} to real time.
This continuation will be further discussed in \cite{highDkink}.   After performing this continuation, one can apply the results directly to the Lorentzian quotients ${\cal M}_{\sigma, T}$.  In doing so, we find as in the Einstein-Hilbert case that the real part of $m$ vanishes, so that the imaginary part of the Lorentzian action (or the real part of the Euclidean action) is just $-\sigma$.  We may thus write

\begin{equation}
e^{iS({\cal M}_{\sigma,T})} = e^{\sigma} e^{-i E_\xi T},
\end{equation}
where $E_\xi$ is the coefficient of the part of the action that is linear in $T$.    Hamilton-Jacobi theory of course guarantees that this $E_\xi$ can still be interpreted as the Killing energy of the higher-derivative theory associated with the boundary Killing field $\xi_\partial$.  As a result, the argument with higher derivative corrections is isomorphic to that given for Einstein-Hilbert gravity but with $\frac{A}{4G}$ replaced by $\sigma$.  And since the difference $\sigma -\frac{A}{4G}$ is perturbatively small, the effect of the higher derivative terms on the final partition functions is perturbatively small as well.

It might seem that this is the final story since,
as phrased above, this argument seems quite satisfying.  However, there is an important caveat which indicates that more work is needed for a full understanding of such higher derivative corrections.  The point is that the above argument was based on taking the Euclidean (i.e., imaginary time) analysis of \cite{Dong:2019piw} and performing an analytic continuation to real time.  For the stationary metrics that give the constrained saddles above, this continuation gives a real metric  (at least after possible appropriate further analytic continuations analogous to $\Omega_E \rightarrow i \Omega$, which is the inverse of the continuation discussed above for rotating black holes).  Our ${\cal M}_{\sigma, T}$ are thus clearly stationary points with respect variations within the class of analytically continued metrics.     But since the general such metric is complex, it is not immediately clear how this variational principle is related to our goal of taking {\it real} Lorentz-signature metrics as the defining contour for the gravitational path integral.  One would thus like to show corresponding results for a variational principle defined on a class of manifestly real spacetimes like the ones defined via cut-and-paste in section \ref{sec:prelim}.  This will in fact be done in \cite{highDkink} at first order in the imaginary part of the parameter $m$ mentioned above.

Now, the imaginary part of $m$ measures the amount of Lorentz boost associated with parallel transport around the conical singularity.  In a real Lorentz-signature spacetime this is not generally infinitesimal, so the extension to finite ${\rm Im} \ m$ is important.  Although this extension is not yet in hand for general metrics, as noted above the analytic extension of stationary Euclidean metrics (i.e., those with a $U(1)$ symmetry) gives real Lorentz-signature metrics.  This means that the analytic continuation of \cite{Dong:2019piw} {\it does} give a good variational principle within the class of real Lorentz-signature spacetimes with a $U(1)$ symmetry generated by some timelike $\xi$, regardless of the value of $m$.

One thus needs only to extend this definition to an appropriate class of real  Lorentz-signature spacetimes that break this symmetry.  But the results will be independent of the details of this extension since, under a variation with momentum $k$ around the $U(1)$, the variation of any diffeomorphism-invariant action about a $U(1)$-symmetric Lorentz-signature configuration must vanish by symmetry when $k \neq 0$; i.e., any failure of the configuration to be stationary must be due to variations that preserve the $U(1)$ symmetry and thus lie within the class for which we can use analytic continuations of the results of \cite{Dong:2019piw}.  We take this as strong evidence that an appropriate higher-derivative action exists even when the imaginary part of $m$ is not infinitesimal, and that the results will be as stated above.

\section{Discussion}
\label{sec:disc}

The goal of our work above was to argue that gravitational path integrals may be defined by integrating over real Lorentz-signature metrics, so long as one allows a certain class of codimension-2 singularities.  In particular, we argued that this approach can avoid the conformal factor problem of Euclidean gravity.   This idea is in the spirit of earlier suggestions in
\cite{Hartle:2020glw,Schleich:1987fm,Mazur:1989by,Giddings:1989ny,Giddings:1990yj,Marolf:1996gb,Dasgupta:2001ue,Ambjorn:2002gr,Feldbrugge:2017kzv,Feldbrugge:2017fcc,Feldbrugge:2017mbc}, though our codimension-2 singularities are not discussed in these works.
A key point in this argument is that the Lorentz-signature path integral is typically oscillatory, and that oscillatory integrals often converge when interpreted as distributions.

However, the above-mentioned singularities create a subtlety. These singularities are Lorentz-signature analogues of Euclidean conical singularities, which are well-known to give delta-function contributions to the Lorentz-signature Einstein-Hilbert Lagrangian $\sqrt{-g} R$, so that the contributions to the Einstein-Hilbert action is proportional to the areas $A[\tilde \gamma]$ of the singular surfaces $\tilde \gamma$.  The subtlety is that,  as noted previously in \cite{Louko:1995jw,Neiman:2013ap,Colin-Ellerin:2020mva}, for some real metrics the coefficient of proportionality can be imaginary or, more generally, it can be complex.     The $A[\tilde \gamma]$ thus define directions within our real Lorentz-signature contour along which the integrand $e^{iS}$ does not oscillate, but instead grows or decays exponentially.

While this may appear to reintroduce an analogue of the Euclidean conformal factor problem, we suggested that it can be controlled by taking the integrals over the areas $A[\tilde \gamma]$ to be performed last, after all of the oscillatory integrals have been performed.  We then illustrated this approach in the context of gravitational partition functions for various ensembles, where we showed that performing the oscillatory integrals first does indeed lead to an additional factor that controls the potentially-dangerous exponential growth and in fact causes the integrals over $A[\tilde \gamma]$ to converge.

In particular, we described how both canonical and microcanonical partition functions for gravitational systems can be formulated in terms of such Lorentz-signature path integrals.  The same is also true of the grand canonical ensemble defined by fixing an angular velocity $\Omega$ and/or an electric potential $\Phi$.   By performing the relevant integrals in the correct order, we showed that the usual Euclidean black holes with positive specific heat -- or their complex rotating generalizations -- provide saddle points $p$ that contribute with non-zero weight $n_p$ to the semiclassical limit of the path integral over a codimension-2 subcontour of the original contour of integration, and that the same is then true of the remaining two integrals.  As noted at the end of section \ref{sec:submain}, this is not quite the same as showing definitively that such saddles contribute to the final partition functions, but we nevertheless take it to provide strong evidence that our Lorentzian formulation yields the expected results in the semiclassical limit.

In studying these partition functions, we have largely relied on the semiclassical approximation.  As is often the case, we have proceeded by identifying interesting classes of saddles and studying their contributions.  In fact, we found it useful to first perform a certain collection of integrals while holding other integration variables constant.  In that sense, we made substantial use of {\it constrained} saddles.  This is reminiscent of the Euclidean work of \cite{Cotler:2020lxj,Cotler:2021cqa}, and it would be interesting to understand this connection in more detail; see also \cite{Bousso:1999iq,Draper:2022xzl}.

As emphasized in the main text, we have taken care to use Morse theory results to show that our constrained saddles contribute with non-zero weight to the semiclassical limits of the path integral defined over a sub-contour of codimension-2, after which we analyzed the remaining two integrals that define the full partition functions.  However,  this does not necessarily imply either that our constrained saddles are the most dominant, or that other yet-to-be identified constrained saddles make no important contributions.

It is thus important to further expand this analysis in several directions.  Within the context of thermodynamics the reader may note that, for a grand canonical ensemble with fixed $\beta, \Omega, \Phi$, our criterion for a black hole solution to contribute is that it should have positive specific heat.  But on physical grounds we would expect contributions only from black holes that are thermodynamically stable, which is a condition that restricts all three eigenvalues of the Hessian for the free energy, and is thus more restrictive than merely requiring positive specific heat.    This suggests that there are further fixed-area saddles that should be included in  \eqref{eq:beta3} and \eqref{eq:beta4}.  On physical grounds, one might expect these to be quotients of {\it additional} black hole exteriors, which in the uncharged context would mean that we should include exteriors for which $\xi_\partial$ is {\it not} the boundary limit of the horizon-generating Killing field.  In other words, it would be physically natural to include black holes whose angular velocity $\Omega$ does not match that naturally given by the metric on $S^1 \times Y$.  Doing so would alter the structure of the singularity that appears after taking the quotient by $e^{\xi T}$, so that there is no longer a well-defined orthogonal plane at each point of $\tilde \gamma$. We might call the result helical singularity, or perhaps a rotating conical singularity. It would thus be very interesting to further investigate the Einstein-Hilbert action in the context of such helical singularities, and to understand their role in the gravitational path integral more generally.

Of course, it would be even more interesting to understand if our analysis can be extended to the full non-perturbative gravitational path integral.  While in higher dimensions the path integral may fail to be well-defined at the non-perturbative level, there has been great success \cite{Saad:2018bqo} in studying the non-perturbative Euclidean path integral in Jackiw-Teitelboim gravity \cite{JACKIW1985343,TEITELBOIM198341}.  It would thus be natural to attempt a similar non-perturbative study of the Lorentzian Jackiw-Teitelboim path integral, both in the thermodynamic context described here and more generally.

A perhaps less ambitious goal is to use a semiclassical approach like the one described here, but to investigate other classes of path integrals.  Examples of particular physical interest include path integrals that compute gravitational Renyi entropies, for which the saddles of interest are replica wormholes.  Forthcoming work \cite{JTRenyi} will study such Lorentzian path integrals in JT gravity, finding a structure very similar to that described here and arguing that the real-time replica wormholes of \cite{Colin-Ellerin:2020mva,Colin-Ellerin:2021jev} contribute to the semiclassical approximation with non-zero weight.  One can also generalize the procedure to consider path integrals that compute states instead of numbers, for example cutting open the construction described here to give a Lorentzian path integral construction of the thermofield-double state.

The important role played by singular spacetimes in our analysis makes it important to fully understand the impact of potential higher-derivative corrections or other issues associated with any UV completion of the low-energy gravitational dynamics.  The arguments presented in section \ref{sec:hd} suggest that, despite the appearance of high curvatures, the effect of such corrections on our partition functions can remain perturbatively small.  However, as also discussed in section \ref{sec:hd}, further work will be required to develop a full framework in which this can be shown to be the case.

Returning now to the context of thermodynamics, recall that we found contributions from the usual saddles that are familiar from Euclidean analyses.  In this sense, we may say that we have derived the Euclidean path integral results from a Lorentzian starting point.  However, there remains an important gap in this circle of ideas.  In particular, due to the conformal factor problem, a well-defined ``Euclidean'' path integral must in fact integrate over some non-trivial contour in the complex plane.  Euclidean approaches thus begin by simply positing such a contour (as in \cite{Gibbons:1978ac}, or more generally as in \cite{Marolf:2022ntb}), which provides an ad hoc element in such arguments.  Closing the circle of ideas completely would thus require using the Lorentzian path integral to directly derive such contour prescriptions for the Euclidean path integral in a general setting, rather than simply verifying their predictions in particular cases.  Doing so would provide a full resolution to the conformal factor problem, perhaps with important implications for cosmology \cite{Feldbrugge:2017kzv,Feldbrugge:2017fcc,Feldbrugge:2017mbc}\cite{DiazDorronsoro:2017hti,Brown:2017wpl,Loges:2022nuw}.   It would also be interesting to understand how such a derivation relates to the non-perturbative Wick rotation described in \cite{Dasgupta:2001ue}.

%%%%%%%%%%%%%%%%%%%%%%%%%
\paragraph{Acknowledgments}
%%%%%%%%%%%%%%%%%%%%%%%%%
It is a pleasure to thank Raghu Mahajan and Jorge Santos for discussions during the initial phase of this work. I also thank Sean Colin-Ellerin, Xi Dong, Henry Maxfield, Mukund Rangamani, Zhencheng Wang for many related discussions.  This work was supported by NSF grants PHY-1801805 and PHY-2107939,  and by funds from the University of California.

\appendix

\section{The spacetimes ${\cal M}_{T,E_\xi}$ are fixed-area saddles}
\label{sec:FAS}

This appendix verifies that the ${\cal M}_{T,E_\xi}$ of section \ref{sec:main} are indeed fixed-area saddles. To do so, we write the Einstein-Hilbert action in the form \eqref{eq:EH}. For variations with $\delta A[\tilde \gamma]=0$ we may ignore the final term in \eqref{eq:EH} and focus on the first two terms.  We should also include any cosmological term or terms describing minimally-coupled matter.  But these will depend on at most first derivatives of the fields.  As a result, while these may be discontinuous at any sewing surfaces $\Sigma_{i\pm}$, they contain no delta-functions and can yield no special contributions near $\tilde \gamma$. We may thus also write the matter and cosmological terms as limits as $\epsilon \rightarrow 0$ of integrals over the region outside $\mathcal{U}_\epsilon$, so that we have formulated our entire variational problem in terms of the $\epsilon \rightarrow 0$ limit of a system with a codimension-1 boundary $\partial \mathcal{U}_\epsilon$.

This is a familiar setting for the computations of variations.  Let us use $S_\epsilon$ to denote the collection of the above terms evaluated at some finite $\epsilon$.  Since the equations of motion are satisfied at each point in the domain of integration, a general variation gives \cite{Brown:1992br}
\begin{equation}
\label{eq:Sepvar}
\delta S_\epsilon = \int_{\partial \mathcal{U}_\epsilon} \sqrt{|h|} \Pi^{ij} \delta h_{ij} + \int_{\partial \mathcal{U}_\epsilon} \sqrt{|h|} \pi_{I} \delta \phi_{I},
\end{equation}
where $i,j$ are coordinates on $\partial \mathcal{U}_\epsilon$, $h_{ij}$ is the induced metric on this surface, $\Pi^{ij} = \frac{1}{8\pi G}\left( K^{ij} - K h^{ij}\right)$ is the Brown-York boundary stress tensor, and $\pi_I$ are the analogous objects for the matter fields.

The notation suggests that we have chosen to write the matter terms with boundary terms at $\partial \mathcal{U}_\epsilon$ appropriate to Dirichlet boundary conditions for the matter fields.  Other choices are also possible.  Because the fields in these terms contain at most first derivatives, such fields can be bounded uniformly in $\epsilon$.  As a result, the fact that the volume $\int_{\partial \mathcal{U}_\epsilon} \sqrt{|h|}$ vanishes at small $\epsilon$ means that any matter boundary terms vanish as $\epsilon \rightarrow 0$.  Thus all possible choices of boundary terms are equivalent as $\epsilon \rightarrow 0$.  Indeed, the same argument shows that we may ignore the matter terms in \eqref{eq:Sepvar} in this limit.

We are thus left with the Brown-York term.  This term can be evaluated using the universal asymptotic form of the metric near a smooth bifurcate Killing horizon,
\begin{equation}
\label{eq:asymptg}
ds^2 \approx - \kappa^2 x^2 d\eta^2  + dx^2 + q_{\alpha \beta} dy^\alpha dy^\beta,
\end{equation}
 expressed in terms of the line element $q_{\alpha \beta} dy^\alpha dy^\beta$ on the bifurcation surface, the Killing parameter $\eta$ along the orbits of $\xi$, the corresponding surface gravity $\kappa$, and the geodesic proper distance $x$ from the bifurcation surface.  Writing the asymptotic metric in the form \eqref{eq:asymptg} also makes use of the fact that the horizon-generating Killing field $\xi$ becomes asymptotically  hypersurface-orthogonal near the bifurcation surface to take $\eta$ constant on such surfaces.  Taking $\partial{\mathcal U}_\epsilon$ to be the surface $x=\epsilon$ one finds
 \begin{equation}
K_{ij} dx^idx^j \approx - \kappa^2 x d \eta^2 \ \ \ {\rm and} \ \ \ \Pi_{ij} \delta h^{ij} \sqrt{q} \approx \frac{1}{x} \frac{1}{8\pi G} q_{\alpha \beta} \delta q^{\alpha \beta} \sqrt{q}= \frac{1}{x} \frac{1}{4\pi G} \delta \sqrt{q}.
\end{equation}
 Inserting this into \eqref{eq:Sepvar} and using $\sqrt{|h|} \approx \kappa x  \sqrt{q}$ from \eqref{eq:asymptg} gives
\begin{equation}
\label{eq:Sepvar2}
\delta S_\epsilon \approx \frac{\kappa T}{2\pi} \frac{1}{4 G}  \int_{\tilde \gamma} \delta \sqrt{q} =  \frac{\kappa T}{2\pi} \frac{1}{4 G} \delta A[\tilde \gamma].
\end{equation}
As a result, so long as $\delta A[\tilde \gamma]=0$, we do indeed find $\delta S_\epsilon \rightarrow 0$ as $\epsilon \rightarrow 0$.  This establishes that ${\cal M}_T$ is a fixed-area saddle as claimed in the main text.
\bibliographystyle{jhep}
	\cleardoublepage

\renewcommand*{\bibname}{References}

\bibliography{negative}

\end{document}